# CEPC Cost Model Study and Circumference Optimization*

Dou Wang#, Jie Gao, Manqi Ruan, Yuhui Li, Haocheng Xu, Yudong Liu, Meng Li, Yuan Zhang, Yiwei Wang, Jiyuan Zhai, Zusheng Zhou

*IHEP, Beijing, China*

**Abstract**

The CEPC is a proposed high luminosity Higgs/Z factory, with the potential to be upgraded to top factory at center-of-mass energy of 360GeV. We perform an optimization study on the circumference of CEPC. We calculate the instant luminosity, the construction and operation cost for different circumferences. With respect to the total cost and average cost per particle, we conclude that the optimal circumference for the CEPC Higgs operation is 80 km. Taking into account of the Z pole operation, the potential high-energy upgrade of CEPC (top factory), the optimal circumference increased to 100 km. The long future proton-proton upgrade of CEPC (SPPC) also favors a larger circumference, and we conclude that 100 km is the global optimized circumference for this facility.

## 1. Introduction

After the discovery of the Higgs boson [1, 2], precision measurement of the Higgs boson properties became a new, and extremely sensitive probe to the new Physics Principles underneath the Standard Model. The Higgs factories based on electron-positron colliders provide crucial information on top of the intensive Higgs program at the Large Hadron Collider (LHC) [3] and its high luminosity upgrade [4-6]. They could significantly boost the precisions of the Higgs property measurements and are regarded as the highest priority for future collider facilities. Multiple electron-positron Higgs factories are proposed, including both linear colliders (International Linear Collider (ILC) [7] and Compact Linear Collider (CLIC) [8]) and circular colliders (Circular Electron Positron Collider (CEPC) [9] and Future Circular Collider (FCC) [10, 11]). Accordingly, intensive physics studies and critical R&D are underway.

The modest Higgs mass of ~125 GeV enables a circular electron-positron collider as a Higgs factory, which has the advantage of a higher luminosity-to-cost ratio compared to the linear collider and the potential to be upgraded to a proton-proton collider to achieve unprecedented high energy (~100 TeV) and discover new physics beyond standard model. Both FCC-ee and CEPC which has a similar scope is a good candidate for the future Higgs factory based on a circular electron-positron collider. The CEPC will operate in three different modes: H ($e^+e^- \rightarrow ZH$), Z ($e^+e^- \rightarrow Z$) and W ($e^+e^- \rightarrow W^+W^-$). The center-of-mass energies are 240 GeV, 91 GeV and 160 GeV, and the luminosities are expected to be $5.0\times10^{34}$, $1.0\times10^{36}$ and $1.6\times10^{35}$ $cm^{-2}s^{-1}$, respectively [9]. A tentative "7-2-1" operation plan is to run CEPC first as a Higgs factory for 7 years and create two million Higgs particles or more, followed by 2 years of operation as a Super Z factory to create

* Supported by the Key Research Program of Frontier Sciences, CAS (Grant No. QYZDJ-SSW-SLH004) and the National Foundation of Natural Sciences (Grant No. 12175249).

# wangd93@ihep.ac.cn

one trillion Z bosons and then 1 year as a W factory to create about 100 million W bosons. After that, the energy of CEPC will be increased to 360GeV with an upgrade, in order to improve the width measurement accuracy of Higgs and increase the accuracy of top mass measurement.

In September 2012, Chinese scientists proposed a 240 GeV Circular Electron Positron Collider (CEPC) with a circumference of 50 km to house two large detectors for Higgs studies. The tunnel for such a machine could also house a Super Proton Proton Collider (SPPC) to reach energies beyond the LHC. It was first presented to the International Committee for Future Accelerators (ICFA) at the Workshop "Accelerators for a Higgs factory: Linear vs. Circular" (HF2012) in November 2012 at Fermilab [12]. A Preliminary Conceptual Design Report (Pre-CDR, the White Report) for a 54 km circular collider was published in March 2015 [13], followed by a Progress Report (the Yellow Report) for the 61 km and 100 km design in April 2017 [14]. The Conceptual Design Report (CDR, the Blue Report) published in August 2018 [15] is a summary of the work done by hundreds of scientists and engineers over the past five years. At that time, we chose 100 km to increase the luminosity of CEPC and push the energy potential of SPPC as much as possible. The luminosity of CEPC is the main reason for choosing 100 km scope with a certain criterion for total power consumption. 30MW synchrotron radiation power per beam is an assumption for CEPC with consideration of the grid distribution status, the electricity capability in China and also the operation cost due to power consumption, while this limitation can be increased to 50MW with upgrade. For CEPC, we are still not quite clear now whether 100 km is the optimum. It is time to look at the circumference of CEPC quantitatively and understand the machine with proper rationale. Therefore, this study develops a cost model to evaluate the cost performance of CEPC. The experiences of existing projects [16-19] are kept in mind and are helpful to CEPC cost study.

## 2. CEPC cost model introduction

The total Higgs number can be expressed as:

$$N_{Higgs} = N_{IP} \cdot L_{design} \times 0.8 \times \sigma \left( N_{year} \times month_{physics} \times 30 \times 24 \times 60 \times 60 \right) \tag{1}$$

where $N_{IP}$ is the number of interaction points, $L_{design}$ is the design luminosity per IP, $N_{year}$ is the required years of operation for certain high energy physics, σ is the cross section of certain physics and $month_{physics}$ is the detector operating months of data taking for each year, which is assumed to be 5 for CEPC. In eq. (1), we introduced a luminosity reduction factor of 0.8 to approximate the real luminosity considering the possible accelerator commissioning status.

The total cost of the Higgs factory is composed of five parts which is expressed as follows:

$$\text{Cost}_{total} = \text{Cost}_{machine} + \text{Cost}_{detector} + \text{Cost}_{elect} + \text{Cost}_{repair} + \text{Cost}_{staff} \tag{2}$$

where the first two parts $\text{Cost}_{machine}$ and $\text{Cost}_{detector}$ are the construction costs for the accelerator and detectors, modeled as follows:

$$\text{Cost}_{machine} = \frac{C}{100} \bullet 24(billion) + 6(billion) \tag{3}$$

$$\text{Cost}_{detector} = 2(billion) \times N_{IP} \tag{4}$$

where $C$ is the circumference of CEPC. The construction costs of accelerator $\text{Cost}_{machine}$ is broken down into a fixed part (6 billion RMB) and a variable part which is linear as the circumference. When the circumference is 100km, the construction costs of accelerator will be 30 billion RMB which is same as the cost estimate in CDR [15]. The construction cost of detector is also estimated in CDR stage which is about 2 billion for each, and hence the total cost for the detectors is the product of single detector cost and the number of interaction point ($N_{IP}$).

The last three parts of eq. (2) are modeled by eq. (5) to eq. (7), which are related to the operating time of the machine, and the total operating year $N_{year}$ is given by eq. (1).

$$\text{Cost}_{elect} = P_{SR} \times 10 \times N_{year} \times month_{operation} \times 30 \times 24 \times 0.5 \tag{5}$$

$$\text{Cost}_{repair} = \text{Cost}_{machine} \times 3\% \times N_{year} \tag{6}$$

$$\text{Cost}_{staff} = \left(\text{Cost}_{machine} \times 1\% + 0.1(billion)\right) \times N_{year} \tag{7}$$

Here, $\text{Cost}_{elect}$ is the cost of electricity where $P_{SR}$ is the SR power per beam and the factor 10 is the magnification coefficient for the total power consumption which is dominated by the operation energy. The factor 0.5 in eq. (5) is the electricity cost per kwh in China and $month_{operation}$ is the machine operation months per year which is assumed to be 9 for CEPC. $\text{Cost}_{repair}$ is the cost of daily care and maintenance of the accelerator where the factor of 3% is from the experience of Beijing Electron Positron Collider (BEPC). $\text{Cost}_{staff}$ is the personnel cost where only the people work on the accelerator is considered and the physicists are not included. When the circumference is 100km, the personnel cost per year is 0.4 billion RMB which correspond to one thousand persons involved and 400 thousand RMB for each person.

## 3. CEPC Cost for the Higgs Factory Configuration

### *3.1 Higgs luminosity scan with different circumference*

#### 3.1.1 scaling law for beam-beam parameter limit and luminosity

In e+ e- storage ring colliders, particles are confined in a bunch due to strong quantum excitation and synchrotron damping effects. The position of each particle is random and the state of the particles can be considered as a gas in which the positions of the particles follow statistical laws. Apparently, synchrotron radiation is the main source of heating. Besides, when two bunches undergo collision at an interaction point (IP), every particle in each bunch will feel the deflected electromagnetic field of the opposite bunch and the particles will suffer from additional heating. The larger the particle population $N_e$ is, the stronger this type of heating becomes and the greater the beam emittance becomes. There is a limiting condition beyond which the beam emittance will blow up [20, 21]. Beam-beam studies shown that crab waist scheme can substantially boost the luminosity of existing and future electron-positron colliders [22, 23], so we proposed this concept in CEPC since 2015 [24]. The beam-beam tune shift is introduced to define the frequency change of the betatron oscillation due to collision. Besides the emittance blow-up mechanism for the maximum value of beam-beam tune shift, a crab waist enhancement factor is introduced to estimate the potential of the beam-beam tune shift. The maximum beam-beam tune shift is also called as beam-beam limit and can be expressed by [21]

$$\xi_{y,\max} = \frac{2845}{2\pi}\sqrt{\frac{U_0}{2\gamma E_0 N_{IP}}} \times F_l \tag{8}$$

Where $N_{IP}$ is the number of interaction points (If there are $N_{IP}$ interaction points, the independent heating effects must be added in a statistical manner), $U_0$ is the synchrotron radiation loss per turn and $E_0$ is the beam energy. The crab waist enhancement factor $F_l$ is related to the Pinwinski angle Φ and the vertical beta function at IP, which can be expressed approximately by:

$$F_l = \begin{cases} \frac{1}{28}\sqrt{\frac{\Phi}{\beta_y^*}} & (\Phi >> 1) \\ \frac{1}{28}\frac{\sqrt{\Phi}}{\left(\beta_y^*\right)^{0.6}} & (\Phi \le 1) \end{cases} \tag{9}$$

and the Pinwinski angle Φ is defined as:

$$\Phi = \frac{\sigma_z}{\sigma_x}\tan\theta_h \tag{10}$$

where $\sigma_x$ and $\sigma_z$ are the horizontal beam size and bunch length at IP and $\theta_h$ is the half crossing angle.

The luminosity of a circular collider is expressed by

$$L[cm^{-2}s^{-1}] = 2.17\times 10^{34}\left(1+r\right)\xi_y \frac{eE_0(GeV)N_b N_e}{T_0(s)\beta_y^*(cm)} F_h \tag{11}$$

where $r=\sigma_y/\sigma_x$ is the aspect ratio of the bunch at IP, $T_0$ is the revolution period, $\beta_y^*$ is the value of the vertical beta function at the interaction point, $\xi_y$ is the vertical beam-beam tune shift, and $F_h$ is the luminosity reduction factor due to hour glass effect, expressed as follows [25]

$$F_h = \frac{\beta_y^*}{\sqrt{\pi}\sigma_z}\exp\left(\frac{\beta_y^{*2}}{2\sigma_z^2}\right) K_0\left(\frac{\beta_y^{*2}}{2\sigma_z^2}\right) \tag{12}$$

where $K_0$ is the zero order modified Bessel function of the second kind.

Combining eq. (8) and eq. (11), one finds that

$$L \propto \xi_{y,\max} I_b \propto \frac{\xi_{y,\max}}{U_0} \propto \frac{1}{\sqrt{U_0}} \propto \sim \sqrt{C} \tag{13}$$

Thus, the luminosity is approximately proportional to the square root of the size of the circular collider when the beam-beam tune shift reaches its limit.

### 3.1.2 CEPC parameters choice with different circumference at Higgs energy

A general method has been developed to optimize the parameters of a circular e+e- Higgs factory by using analytical expressions for the maximum beam-beam parameter and the beamstrahlung beam lifetime, starting from a given design goal and technical constraints [24, 26, 27]. A parameter space has been explored. Based on the scan of beam parameters and RF parameters, a set of optimized parameter designs for CEPC with different circumference was proposed. The luminosity for the Higgs energy as a function of the size of the circular collider is shown in Fig. 1. To maximize the luminosity for each circumference, the IP beam parameters and the lattice structure were carefully optimized. For example, the FODO length is 55m for 100km circumference, while it is 80m for 200km circumference. Also, we chose different β* for different circumferences to achieve the maximum beam-beam tune shift, and meanwhile to keep the beam lifetime (dominated by the beamstrahlung lifetime and barbar lifetime) at the same level for different machine sizes. The requirement for the dynamic aperture energy acceptance of larger ring is smaller than the smaller ring according to the beamstahlung lifetime estimation [28, 29], so a slightly smaller $\beta_y$* can be used for a larger ring. Additionally, the RF voltage is lower for a larger ring than for a smaller ring due to the lower momentum compaction factor of the lattice.

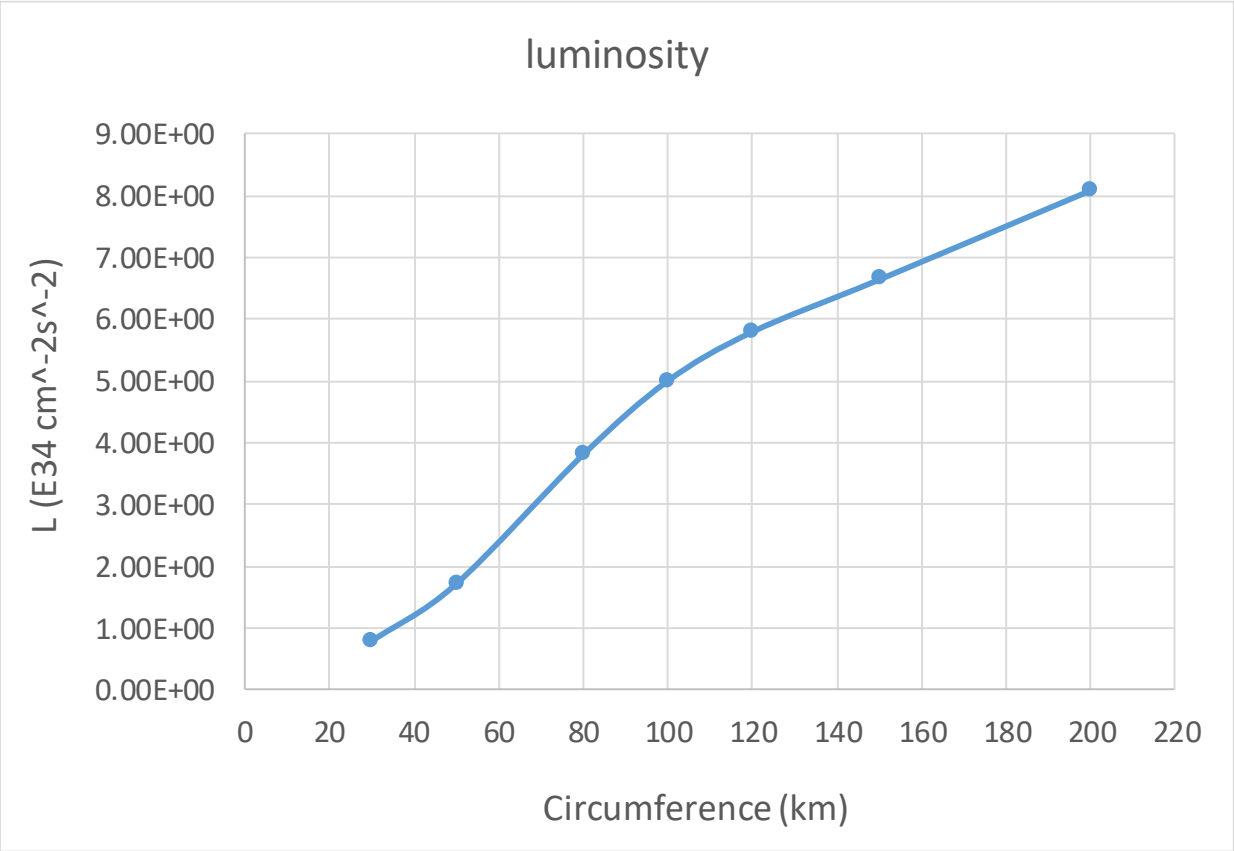

Fig. 1: Results of the luminosity optimization studies for the CEPC collider at the beam energy of 120 GeV (Higgs factory configuration) vs machine circumference ($P_{SR}$=30MW).

### *3.2 CEPC 4IP scheme*

CEPC schematic design with four IPs is also studied without detail optics design. For CEPC, we keep focus on 2 IP scheme during TDR period [9], while FCC-ee has moved on to 4 IP scheme after its CDR publication [30-32]. The layout of CEPC four IP scheme is shown in Fig. 2. We assume equal space between four IPs and shared RF systems are used. The RF stations are located in the middle of two adjacent IPs to ensure that the same beam energy arrives at each detector. Therefore, there are a total of four RF sections in the colliding rings. In the 4 IP scheme, there are two bunch trains per beam and therefore it is a two-by-two collision, which is different from the 2 IP scheme.

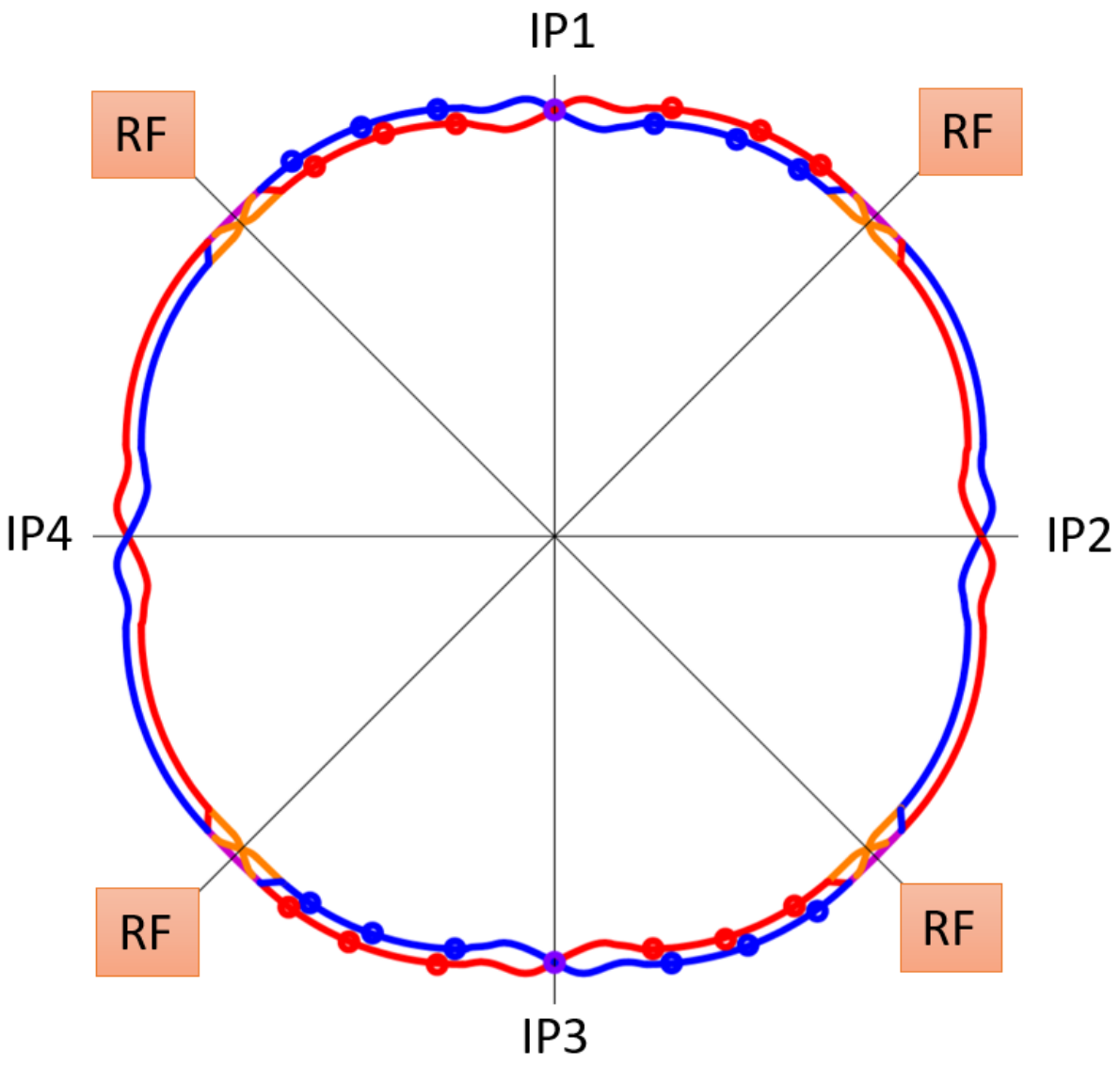

Fig. 2: The layout of CEPC 4 IP scheme (Blue is for the electron ring and red is for the positron ring).

From eq. (8), the maximum beam-beam tune shift is inversely proportional to the square root of the IP number, and from eq. (13), the luminosity per IP is also inversely proportional to the square root of the SR loss $U_0$. Meanwhile considering the total length reduction of arc section due to more insertion space for detectors and beam focusing, $U_0$ will increase slightly and hence the single IP luminosity for different IP number can be scaled by the luminosity of the 2 IP case by:

$$L(N_{IP}) = L(2IP)\frac{\sqrt{\frac{2}{N_{IP}}}}{\sqrt{1+\frac{4*(N_{IP}-2)}{C}}} \tag{14}$$

*3.3 Cost of Higgs factory*

3.3.1 required years of operation

Number of years of required operation can be calculated according to eq. (1) and the luminosity value in Fig. 1, where the cross section at 240GeV is assumed to be 200fb. Fig. 3 shows the operating years for a Higgs factory with different operating conditions and different physics goals. For the operating conditions with 50MW SR beam power, the luminosity per IP can be scaled from 30 MW case linearly with the beam power. And the single IP luminosity for the 4 IP scheme can be scaled from the 2 IP case by eq. (14). From Fig.3, we know that about 1.5 million Higgs particles can be achieved with 7 years' planning running time if we choose a circumference of 100 km and 30 MW operating conditions.

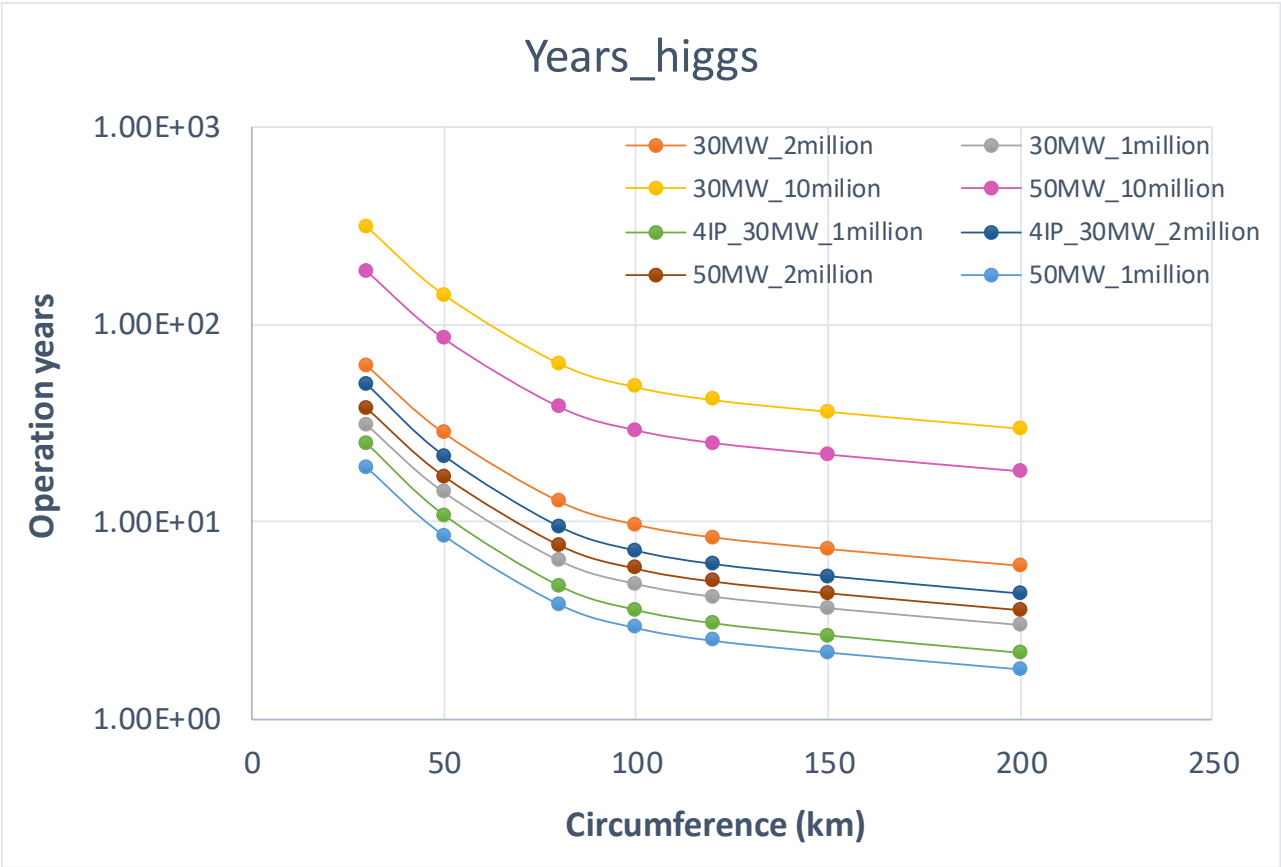

Fig. 3: Required years of operation for the CEPC collider at the beam energy of 120 GeV (Higgs factory configuration) with different operation conditions and different physics goal.

### 3.3.2 total cost for Higgs factory

Fig. 4 shows the CEPC total costs only for a Higgs factory with different operating conditions. The lower three lines are the costs for the 1 million Higgs goal and the upper three lines are the costs for 2 million Higgs goal. The minimum cost for a Higgs factory is RMB 40 billion and the optimum circumference is 80 km. If the requirement of total Higgs bosons reach 2 million, the total cost is almost the same for 80 km and 100 km option.

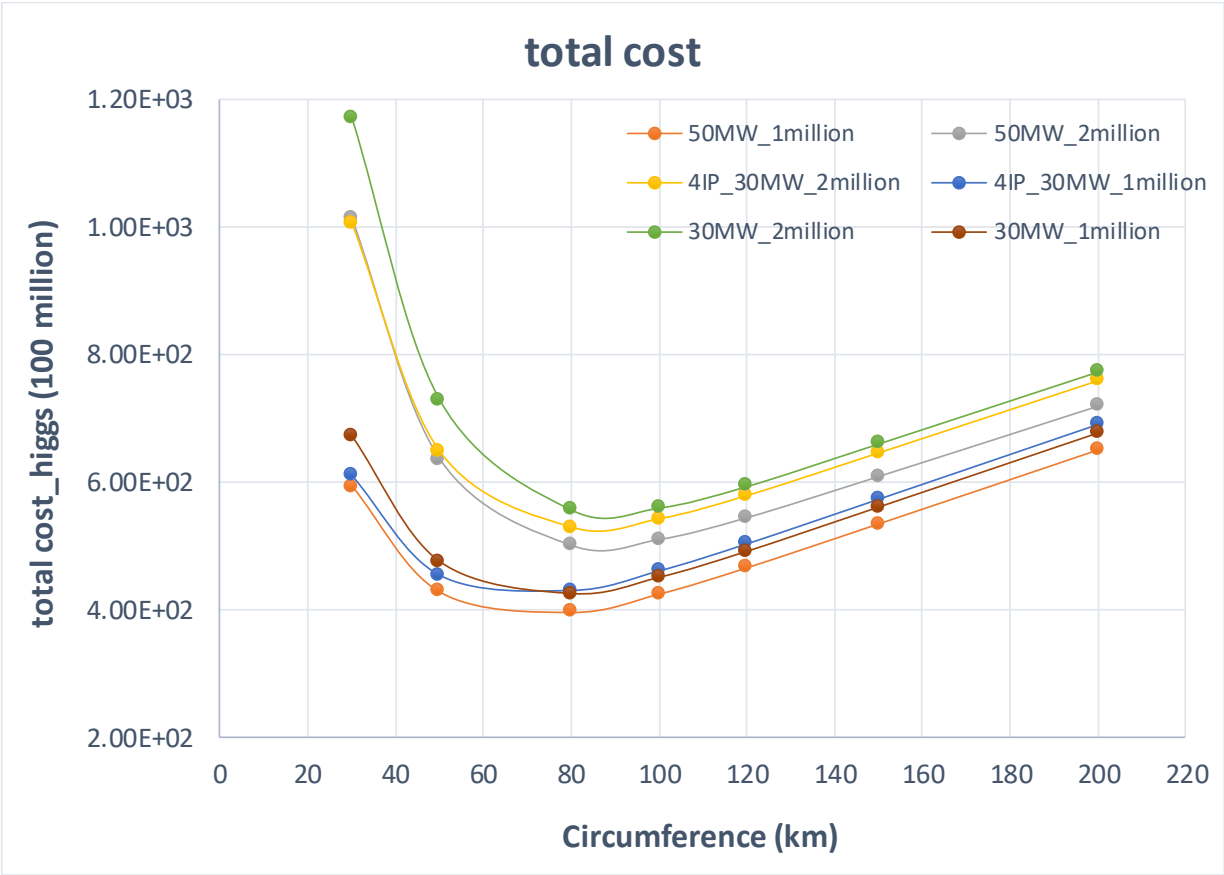

Fig. 4: The total cost for the CEPC collider at the beam energy of 120 GeV (Higgs factory configuration) vs machine circumference.

### 3.3.3 cost per Higgs particle

Fig. 5 shows the cost per Higgs particle produced under different operation conditions. We can see that all lines are distributed in three series, which is related to the physical requirements. The upper there lines are the single Higgs cost for the 1 million Higgs goal, which is roughly 45 thousand RMB at 100 km circumference; the middle three lines are the single Higgs cost for the 2 million Higgs goal which is roughly 28 thousand RMB at 100 km circumference, and the lowest two dashed lines are the single Higgs cost even for the 10 million Higgs goal with much lower Higgs cost which are shown here just as a reference. We know that the single Higgs cost is dominated by the physical requirements rather than the detail operation condition.

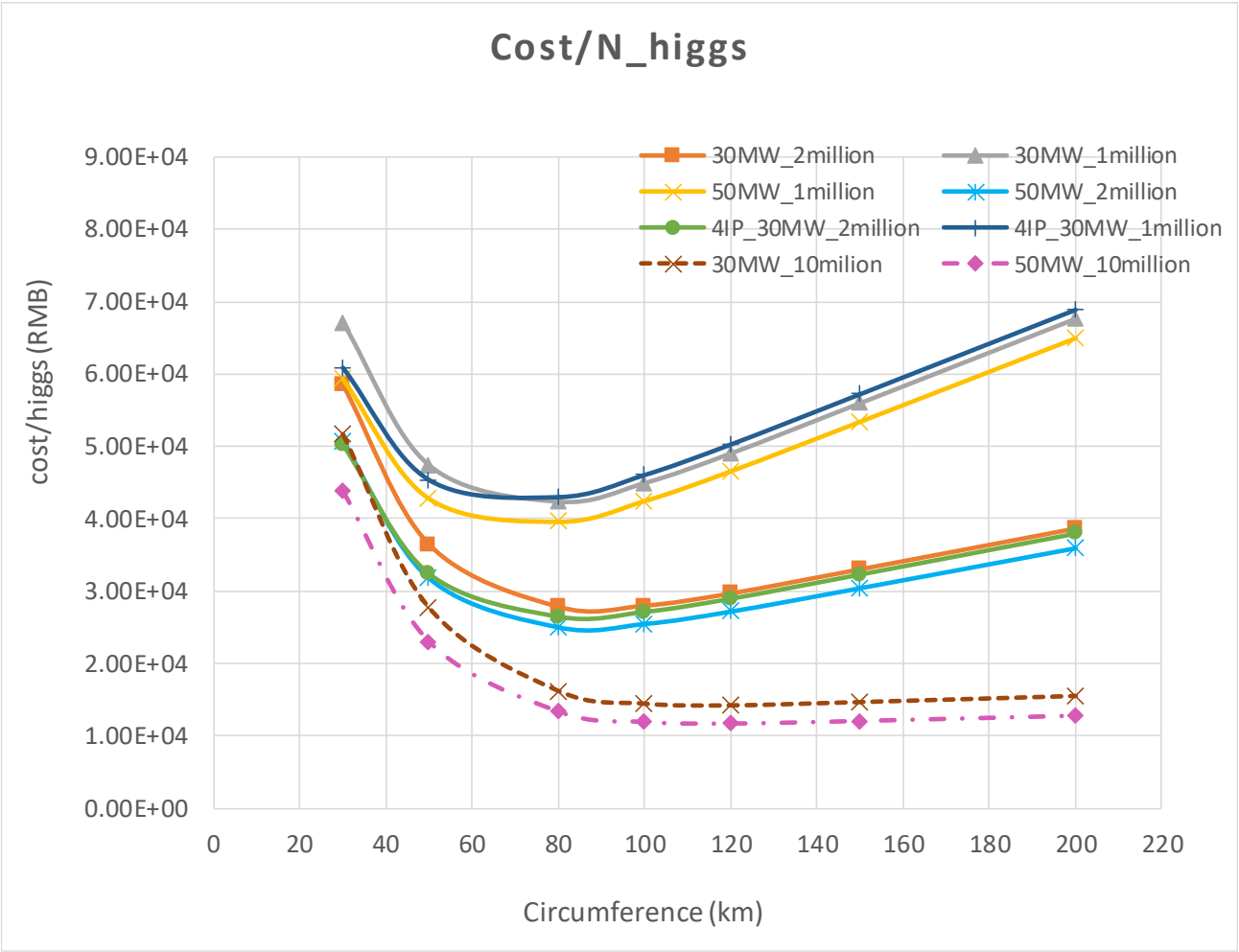

Fig. 5: The cost per Higgs particle for the CEPC collider at the beam energy of 120 GeV (Higgs factory configuration) with different operation conditions and different physics goal.

### 3.3.4 comparison between 2 IP and 4 IP

Fig. 6 shows the total cost comparison of a Higgs factory for the 2 IP and 4 IP schemes with different SR power. For 30MW beam power, the crossing point of 2 IP and 4 IP scheme refer to 1.5 million Higgs which coincides with the real achievement for 7 years of Higgs operation. Thus, the 4 IP scheme is more economical when the requirement for Higgs particles is more than 1.5 million. In the 50 MW case, the crossing point of 2IP and 4IP scheme refer to 1.9 million Higgs which is also close to the real achievement with 7 years of Higgs operation. Overall speaking, the total cost of the Higgs factory is almost same for the 2 IP scheme and the 4 IP scheme, regardless of how much beam power we use when the requirement for Higgs particles is between 1.5 million and 1.9 million.

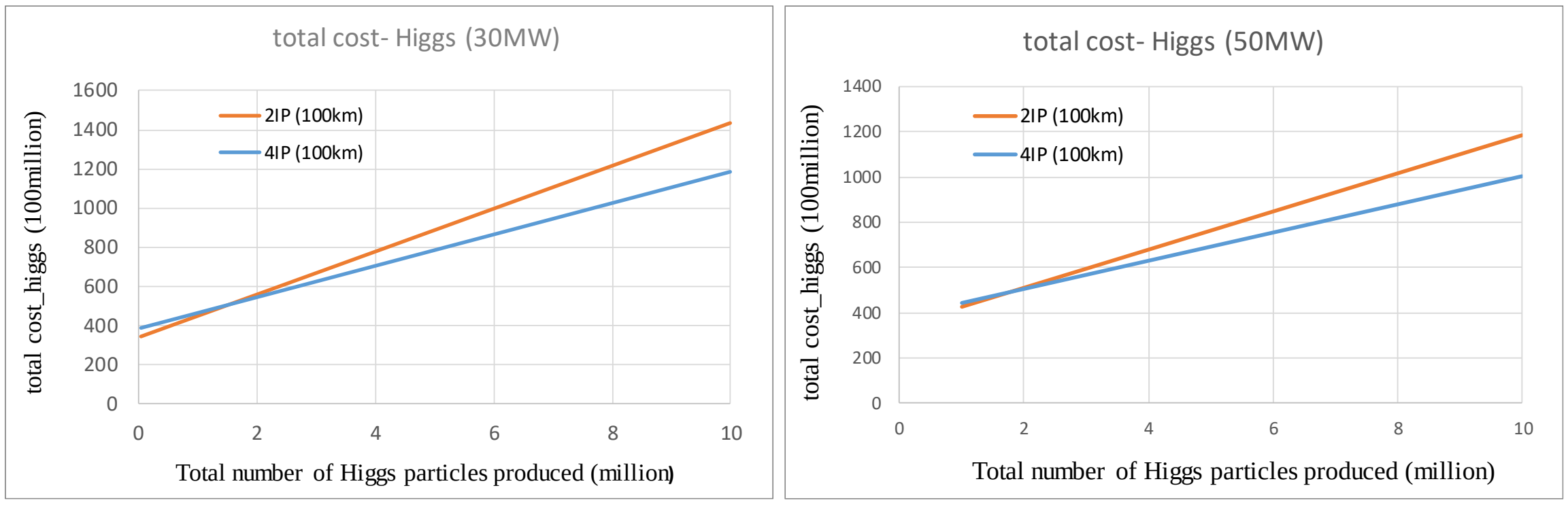

Fig. 6: The total cost comparison of a Higgs factory between 2IP and 4IP scheme with different SR power (left: 30MW, right: 50MW).

## 4. CEPC cost combing Z factory

*4.1 Z luminosity scan with different circumference*

Just as with the Higgs energy, we have optimized the beam parameters and the phase advance of FODO cell while the lattice structure is kept same as the lattice of the Higgs factory configuration. The phase advance of FODO cell is 60 degree while the circumference is smaller than 120km, and the FODO phase advance is reduced to 45 degrees when the circumference is larger than 120 km to achieve the highest luminosity. We have found that larger rings require much more bunches than smaller rings because of the lower beam-beam limit and lower bunch charge. While the bunch number at Z pole is limited by the electron cloud instability and the maximum bunch number for a 200 km long ring is assumed to be 35000 by a rough analytical estimate. The optimized luminosity at the Z pole with different circumferences is shown in Fig. 7 and the corresponding SR power for each beam is shown in Fig. 8. From Fig. 8 we see that the beam power cannot reach the full design value (30 MW) because of the electron cloud instability, so that the luminosity in Fig. 7 starts to decrease after 120 km.

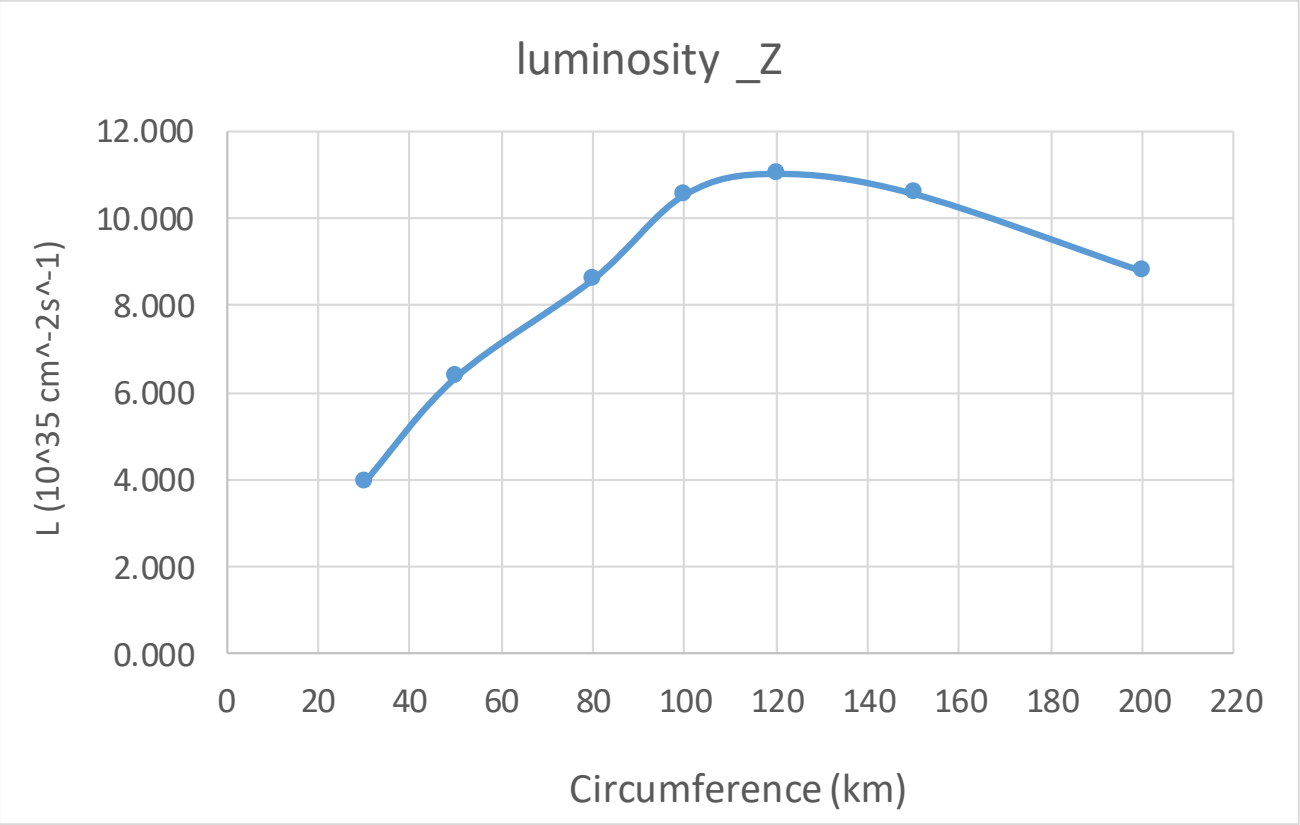

Fig. 7: Results of the luminosity optimization studies for the CEPC collider at the beam energy of 45.4 GeV (Z factory configuration) vs machine circumference.

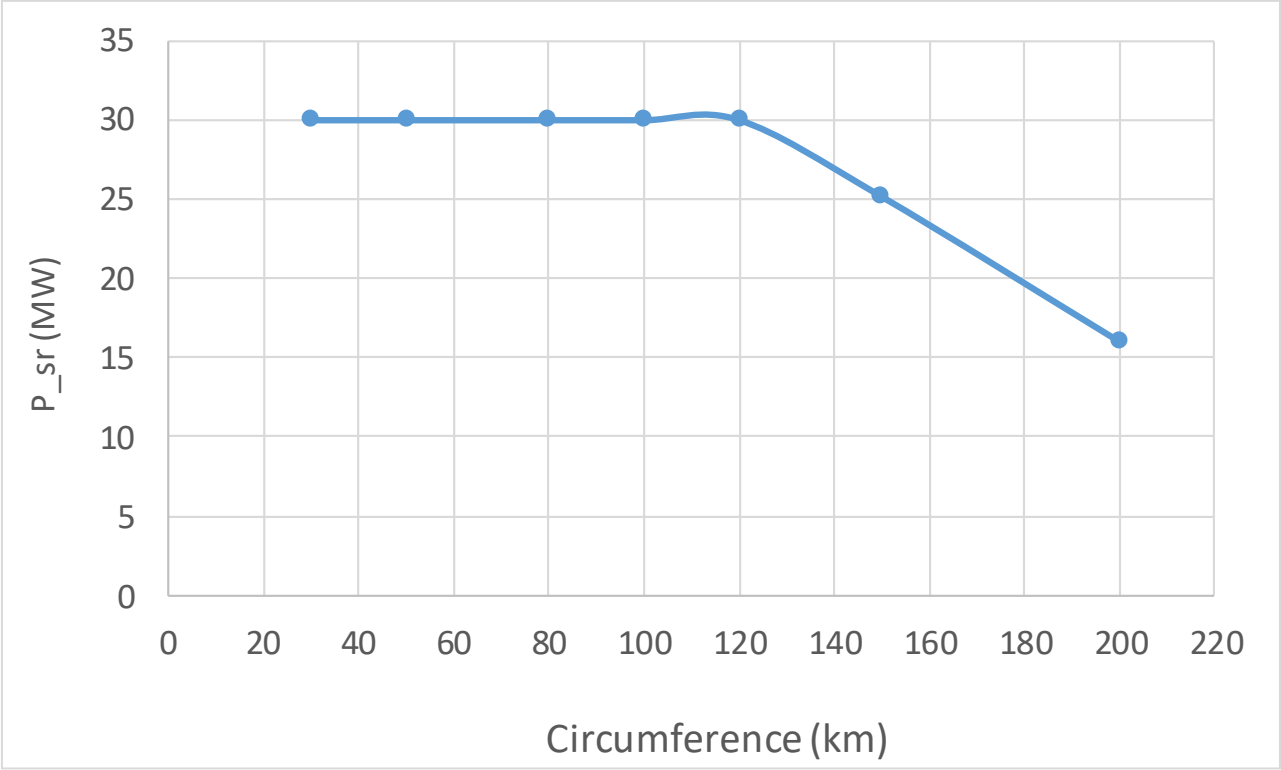

Fig. 8: SR power per beam for the CEPC collider at the beam energy of 45.4 GeV (Z factory configuration) vs machine circumference.

*4.2 CEPC cost considering compatibility with Z factory*

4.2.1 required years of operation for Z factory

The total operating year of the CEPC as a Z-factory can be calculated according to eq. (1) and the luminosity value in Fig. 7 where the cross section at 91GeV is assumed to be 30nb. Fig. 9 shows the operation years for a Z-factory with different physics goals. From Fig. 9, we know that about 1.3 tera of Z bosons can be achieved with 2 years' planning running time if we choose a circumference of 100 km.

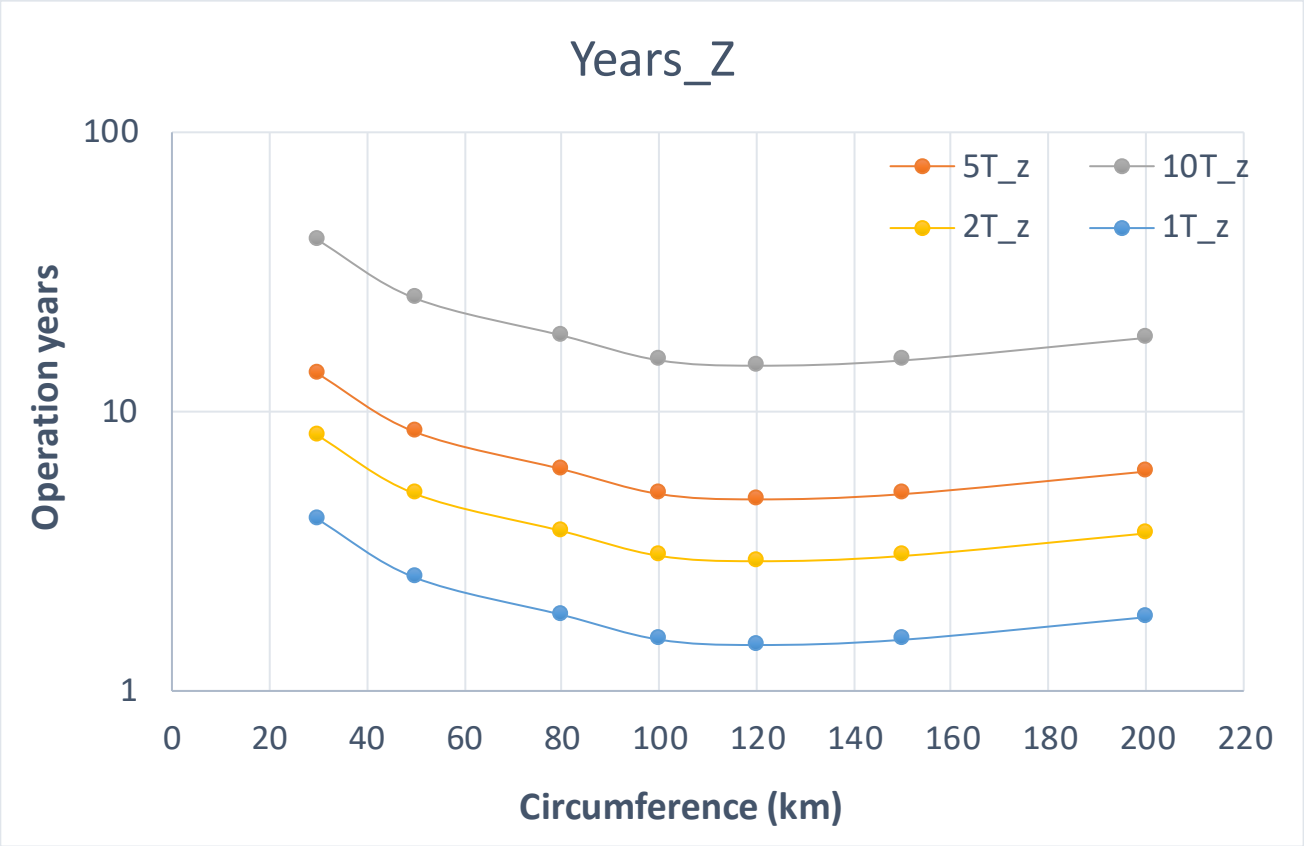

Fig. 9: Required years of operation for the CEPC collider at the beam energy of 45.4 GeV (Z factory configuration) vs machine circumference.

4.2.2 cost per Z particle

At the Z pole, the cost of electricity in eq. (5) should be changed to eq. (15) due to lower energy running.

$$\text{Cost}_{elect} = P_{SR} \times 6 \times N_{year} \times month_{operation} \times 30 \times 24 \times 0.5 \tag{15}$$

Fig. 10 shows the cost per Z boson with different physics goal. We can see that Z boson becomes cheaper if we require more Z bosons and the cost of single Z boson with the goal of 1 tera Z is roughly 0.035 RMB at 100 km circumference. We also know that the Z-factory prefers the smaller ring if we only consider Z physics without Higgs.

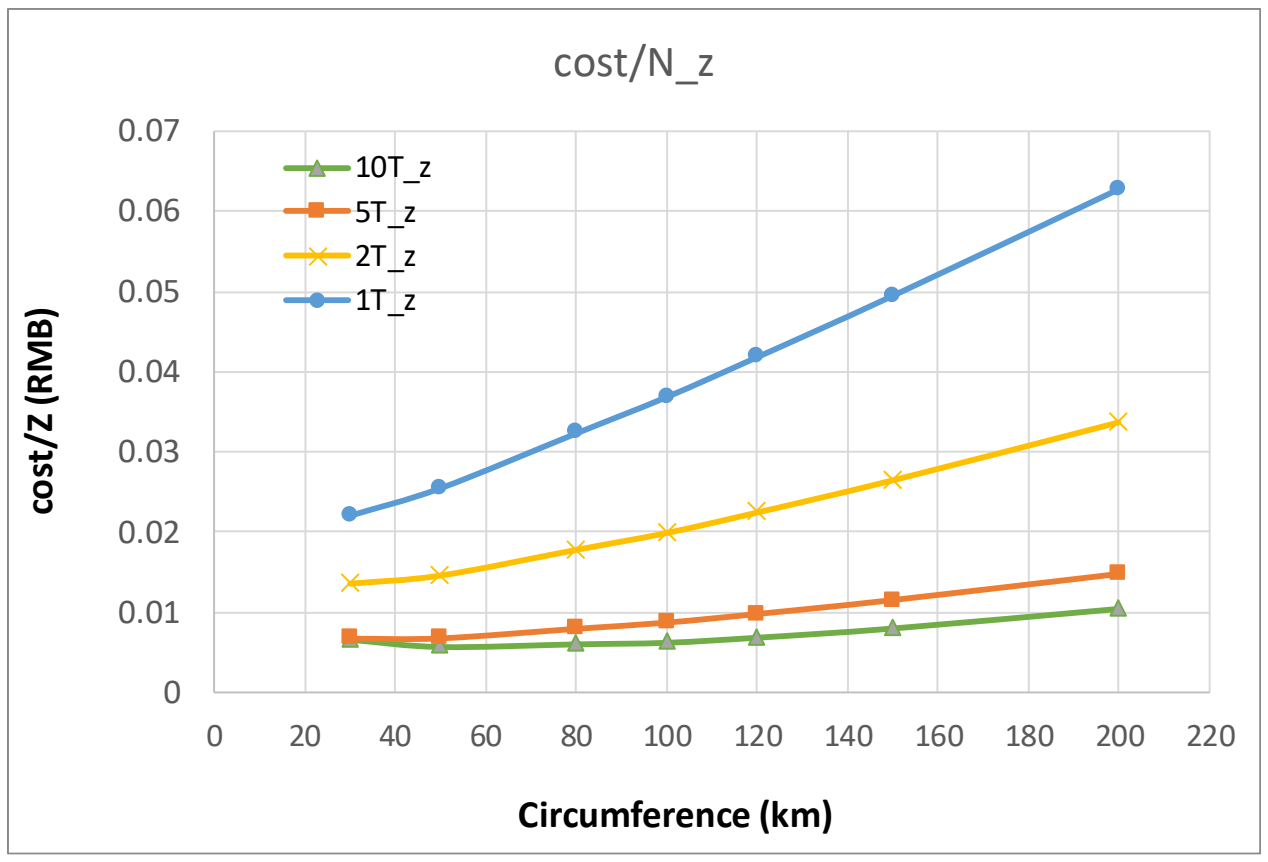

Fig. 10: The cost per Z boson for the CEPC collider at the beam energy of 45.5 GeV (Z factory configuration) with different physics goal.

4.2.3 cost per Higgs particle combining Higgs and Z

The cost of each Higgs boson can be reduced if we combine Higgs physics and Z physics. The cost per Higgs can be revised considering the construction cost allocation between the Higgs factory and the Z factory according to their operation year, which is given by eq. (16) eq. (17). As an example, Fig. 11 shows the cost per Higgs particle for a given operating condition with different Z bosons goal. We can see that the single Higgs cost for the case with 1million Higgs requirement and 30MW SR power can be reduced from 45 thousand RMB to 36 thousand RMB while including 1 tera of Z bosons with 100 km circumference.

$$F_{Higgs} = \frac{Year_{Higgs}}{Year_{Higgs} + Year_Z} \tag{16}$$

$$\text{Cost}_{machine@Higgs} + \text{Cost}_{detector@Higgs} = \left(\text{Cost}_{machine} + \text{Cost}_{detector}\right) \times F_{Higgs} \tag{17}$$

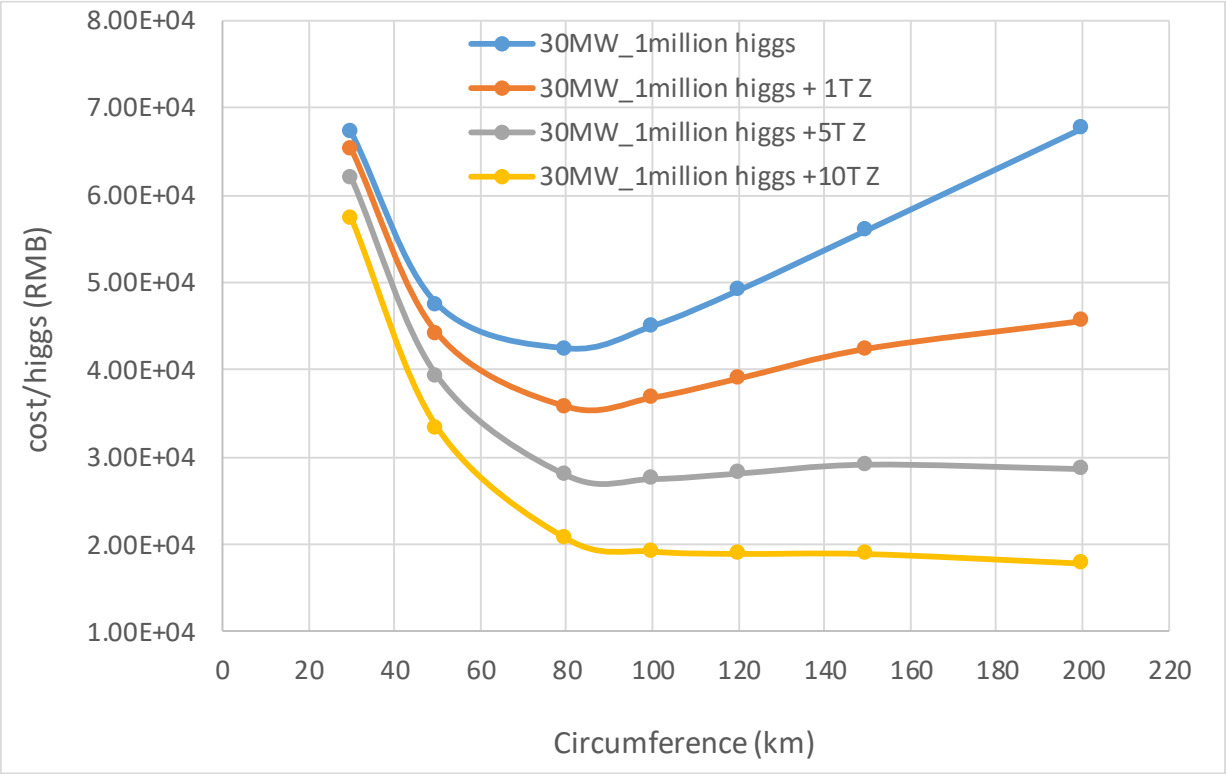

Fig. 11: The cost per Higgs boson for the CEPC collider combing Higgs physics and Z physics vs machine circumference.

4.2.4 CEPC total cost combining Higgs and Z

Fig. 12 shows the CEPC total cost combing Higgs physics and Z physics with different operation conditions. The lower two lines are the costs for 1 million Higgs goal with 1 tera Z bosons and the upper two lines are the costs for 2 million Higgs goal with 1 tera Z bosons. The minimum cost for a CEPC is RMB 42 billion and the optimum circumference is 80 km for the case of 1 million Higgs. Again, if the total demand for Higgs bosons is more than 2 million, the cost of 100 km circumference would be almost same as that of 80 km.

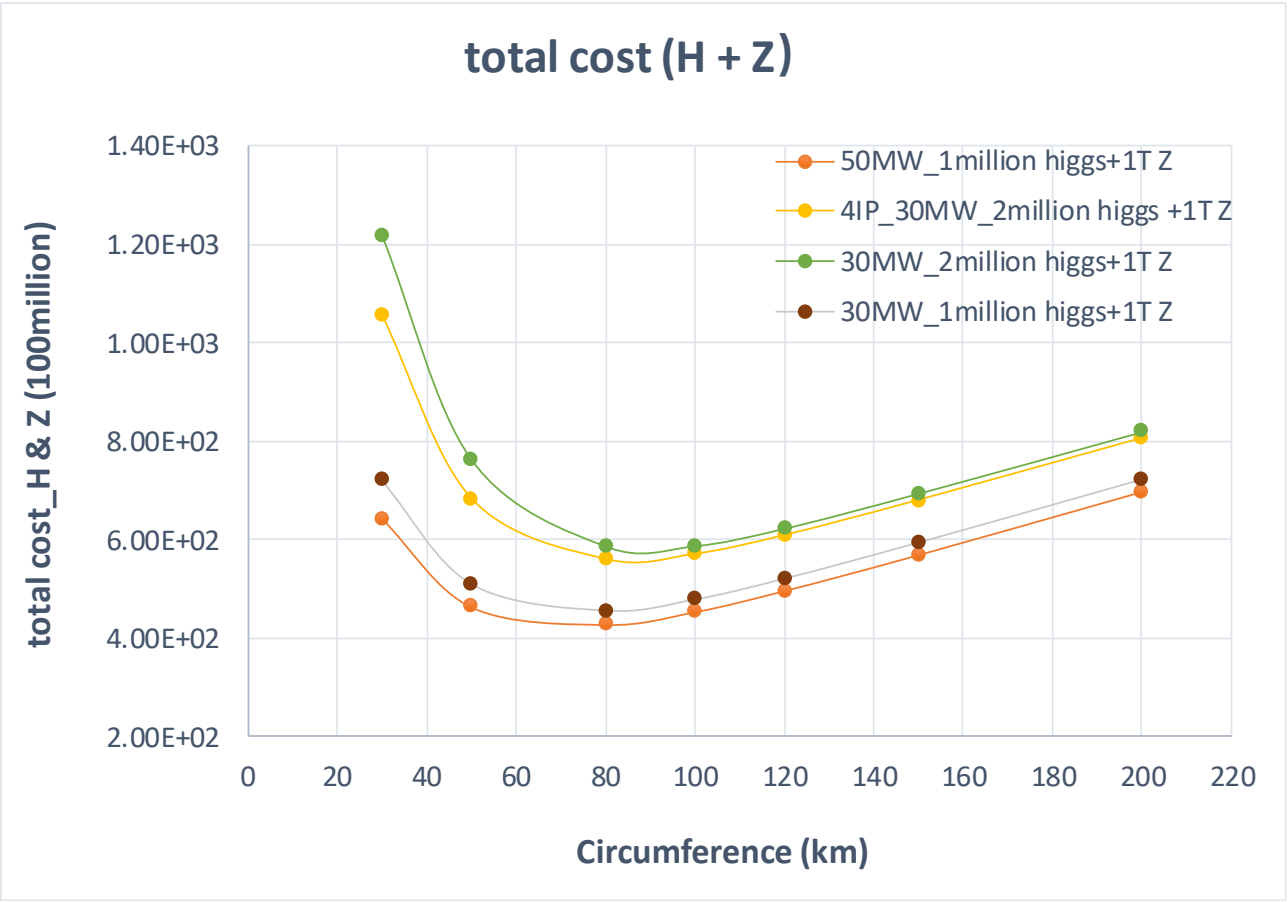

Fig. 12: CEPC total cost combing Higgs physics and Z physics with different operation conditions and different physics goal.

## 5. CEPC cost combing tt physics

### *5.1 tt luminosity scan with different circumference*

Just as with the Higgs energy, the beam parameters at 180 GeV are optimized at each circumference to achieve higher luminosity as much as possible, while the lattice structure is kept same as the lattice of Higgs factory configuration. So far, the CEPC parameters and the lattice design have been optimized at the Higgs energy. For higher energy of tt, if we still use the same lattice as Higgs, the strength of the FD SC quadrupoles will exceed the maximum capability of the technology. Moreover, the beam stay clear region will be larger than the beam pipe designed for Higgs energy because the emittance of ttbar becomes larger than that of Higgs. So for tt mode, we need to redesign the IR lattice and relax the $\beta x^*/\beta y^*$ which can fulfill the constraint for the FD quadrupole strength and beam stay clear region. A larger horizontal $\beta x^*$ (~1 m for 100km) is chosen to ensure that the new beam stay clear region would not be larger than that of Higgs at IR SC quadrupole region, and a larger vertical $\beta y^*$ (~2.7 mm for 100km) is chosen to get larger DA energy acceptance according to stronger beamtrahlung effect with higher energy. Overall speaking, with larger ring, the IP vertical $\beta y^*$ can be slightly lower than smaller ring due to smaller requirement of DA energy acceptance and lower beam-beam limit. In addition, the RF voltage of smaller ring is higher than larger ring due to larger momentum compaction factor, and hence the smaller collider under 50 km cannot work at tt energy because the space ratio kept for RF systems would be too high compared with the total circumference. The optimized luminosity at tt energy with different circumferences is shown in Fig. 13 with 30MW SR power for each beam.

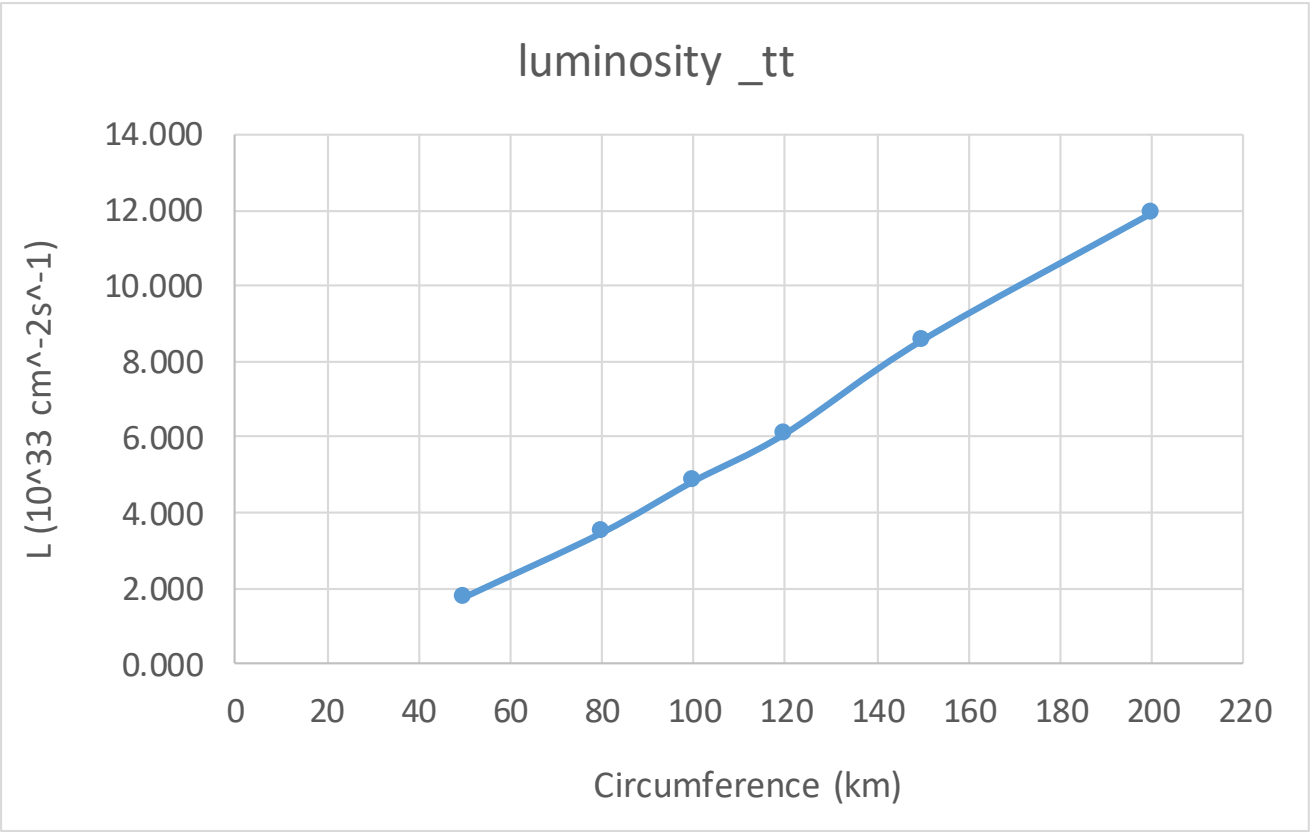

Fig. 13: Results of the luminosity optimization studies for the CEPC collider at the beam energy of 180 GeV vs machine circumference ($P_{SR}$=30 MW).

## *5.2 CEPC cost considering compatibility with top factory*

### 5.2.1 required years of operation for top factory

The total operation year of CEPC at tt energy can be calculated according to eq. (1) and the luminosity value in Fig. 13 where the cross section at 360GeV is assumed to be 500fb. Also considering two top quarks will be produced by one reaction, fig. 14 shows the operation years for the tt study with different physics goals. From fig. 14, we know that the minimum operation year for 1 million top quarks is about 6 years if we choose 100 km circumference and 50MW running condition.

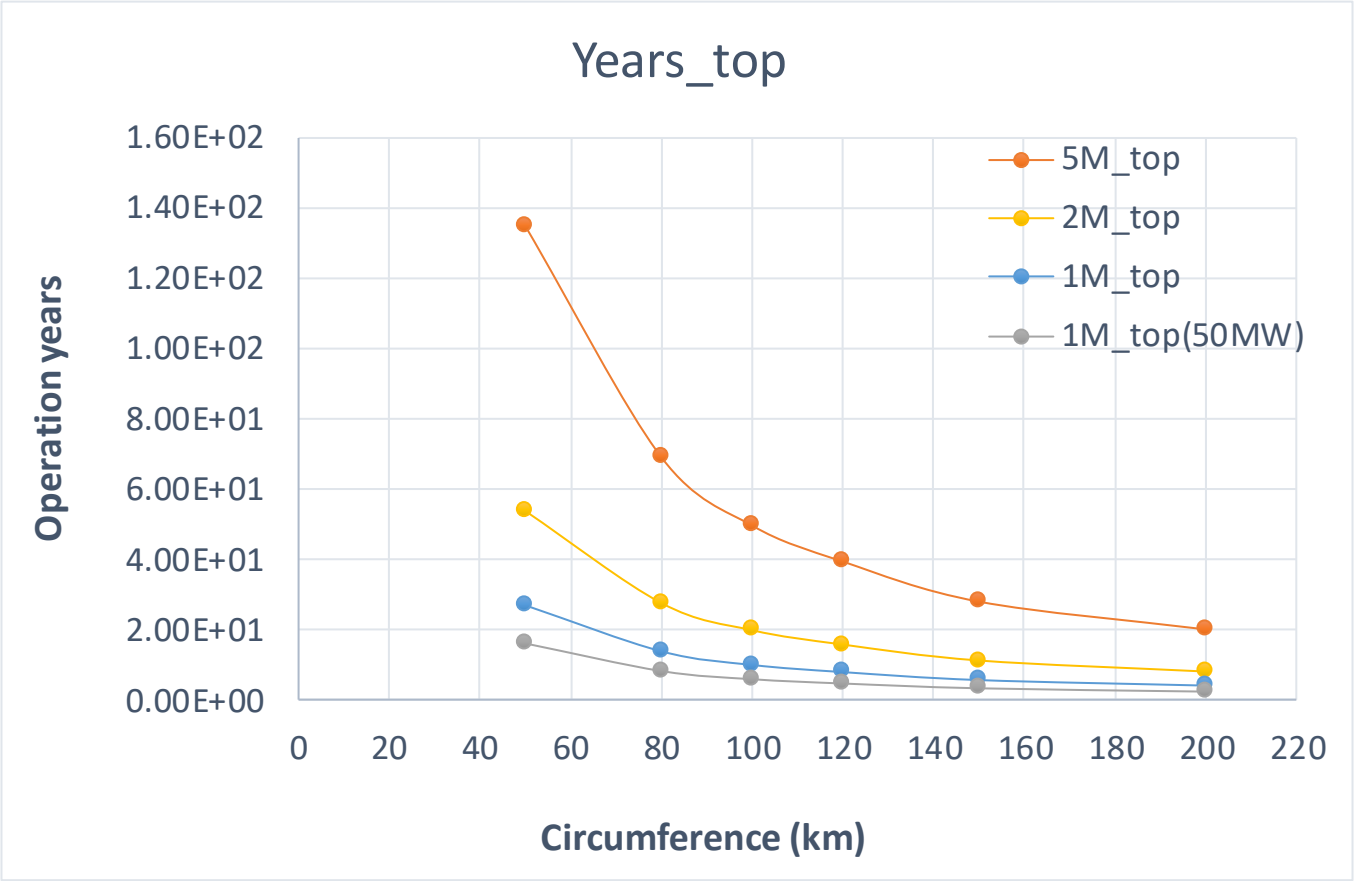

Fig. 14: Required years of operation for the CEPC collider at the beam energy of 180 GeV vs machine circumference.

### 5.2.2 cost per top quark

For the tt physics study, the construction cost for the accelerator in eq. (3) should be modified to eq. (18) taking into account the upgrade cost for the additional RF systems at tt energy. The machine cost will increase due to the upgrade of RF system because tt operation needs much more RF cavities. In eq. (18), the factor $k$ is related to the detail RF voltage and its value is 1 when the circumference is 100 km. The cost of electricity in eq. (5) should be modified to eq. (19) due to the higher beam energy. Fig. 15 shows the cost per top quark with different physics goal. We can see that each top quark becomes cheaper if more top quarks are expected. Also the highest cost of each top quark for 1 million total top quarks and 30MW SR power is roughly 68 thousand RMB at 100 km circumference. Furthermore, we can see that the tt physics favors the larger ring if we only consider the tt physics compared to the Higgs energy.

$$\text{Cost}(machine) = \frac{C}{100} \bullet 24(billion) + 6(billion) + k \times 6(billion) \tag{18}$$

$$\text{Cost}(elect) = P_{SR} \times 13 \times year \times month_{operation} \times 30 \times 24 \times 0.5 \tag{19}$$

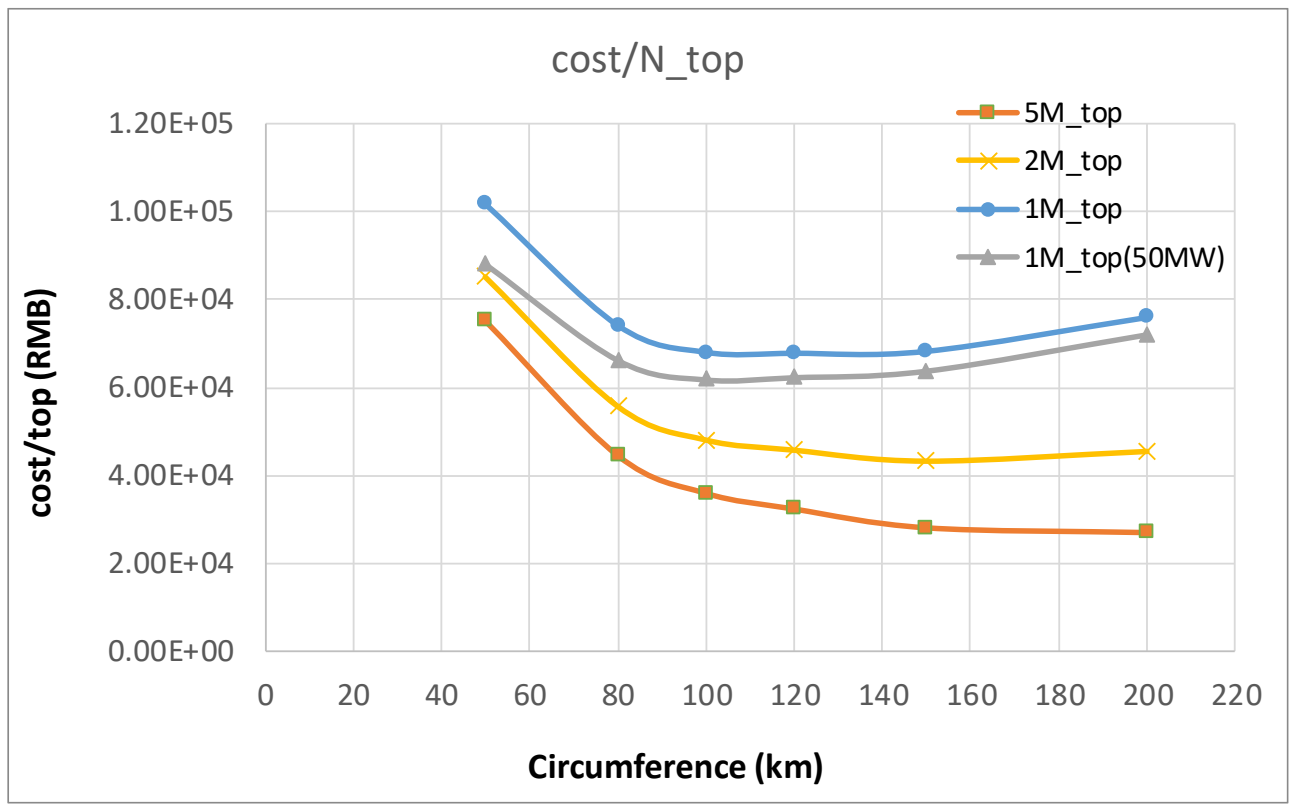

Fig. 15: The cost per top quark for the CEPC collider at the beam energy of 180 GeV with different physics goal vs machine circumference.

### 5.2.3 CEPC total cost combining Higgs, Z and tt operation

Fig. 16 shows the total costs of the CEPC for the combination of Higgs physics, Z physics and top quark under different operating conditions. The lower four lines are the costs without top quark study and the upper four lines are the costs with 1 million top quarks. The minimum CEPC cost is about RMB 73 billion if we consider Higgs, Z and top quark physics together and the optimal circumference changes to 100km.

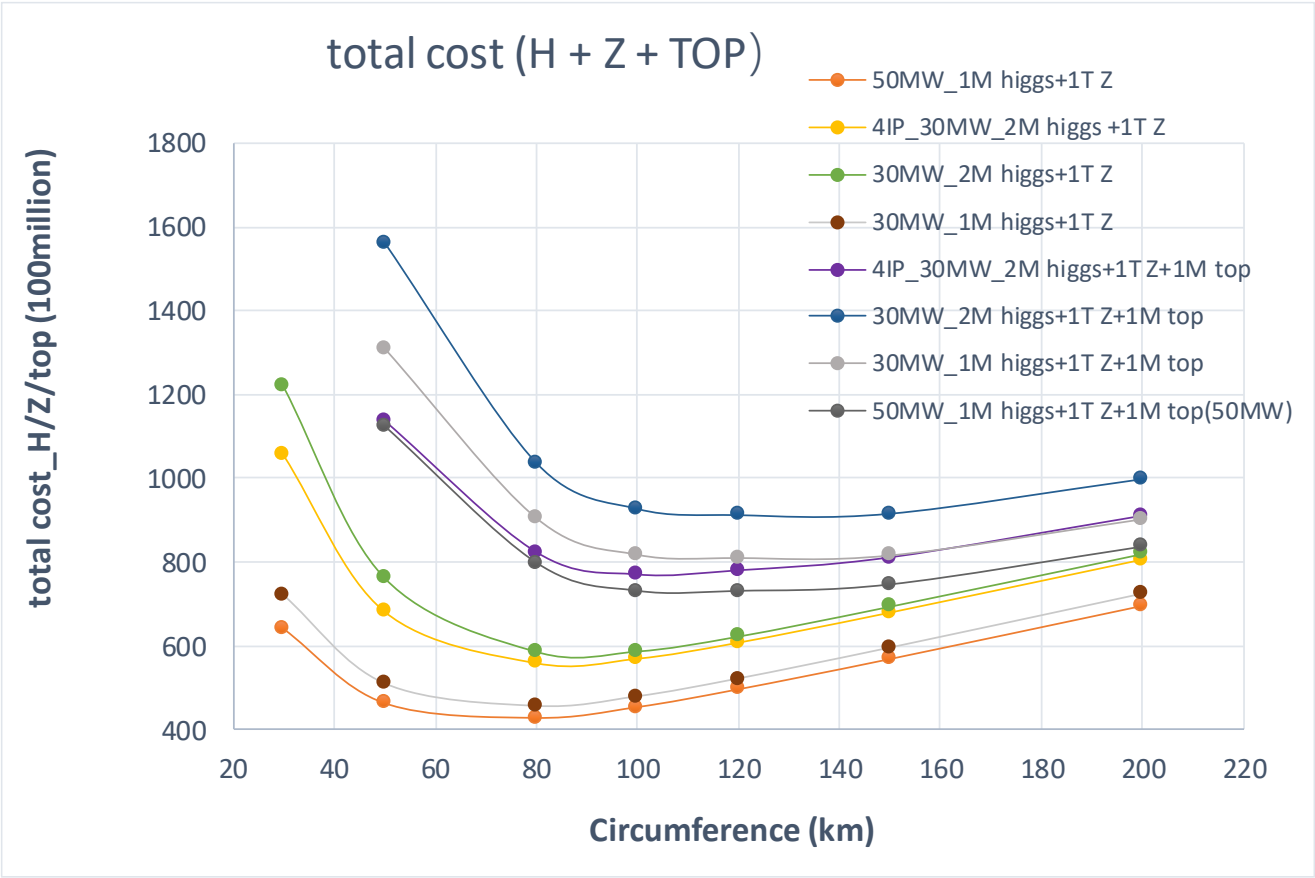

Fig. 16: CEPC total cost combing Higgs physics, Z physics and top quark with different operation conditions and different physics goal vs machine circumference.

## 6. Energy potential of SPPC

The CEPC has the potential to be upgraded to a proton-proton collider to achieve unprecedented high energy and discover New Physics. It is planned to build a super proton-proton collider (SPPC) in the same tunnel of the CEPC after 10 years of operation. The construction of CEPC and SPPC in a common accelerator complex also provides a great opportunity to realize collisions of ultra-high energy protons or ions with very high energy electrons or positrons (e-p or e-A). Fig. 17 shows the dipole strength of SPPC related with the different circumference and Fig. 18 shows the center-of-mass energy of SPPC related with the different circumference. The energy potential of SPPC is strongly depends on the size of the collider and the technology of the superconducting dipole magnets. From Fig. 18, it can be seen that a larger ring can achieve a higher energy with a certain technology level and that a larger circumference is helpful to push the energy frontier.

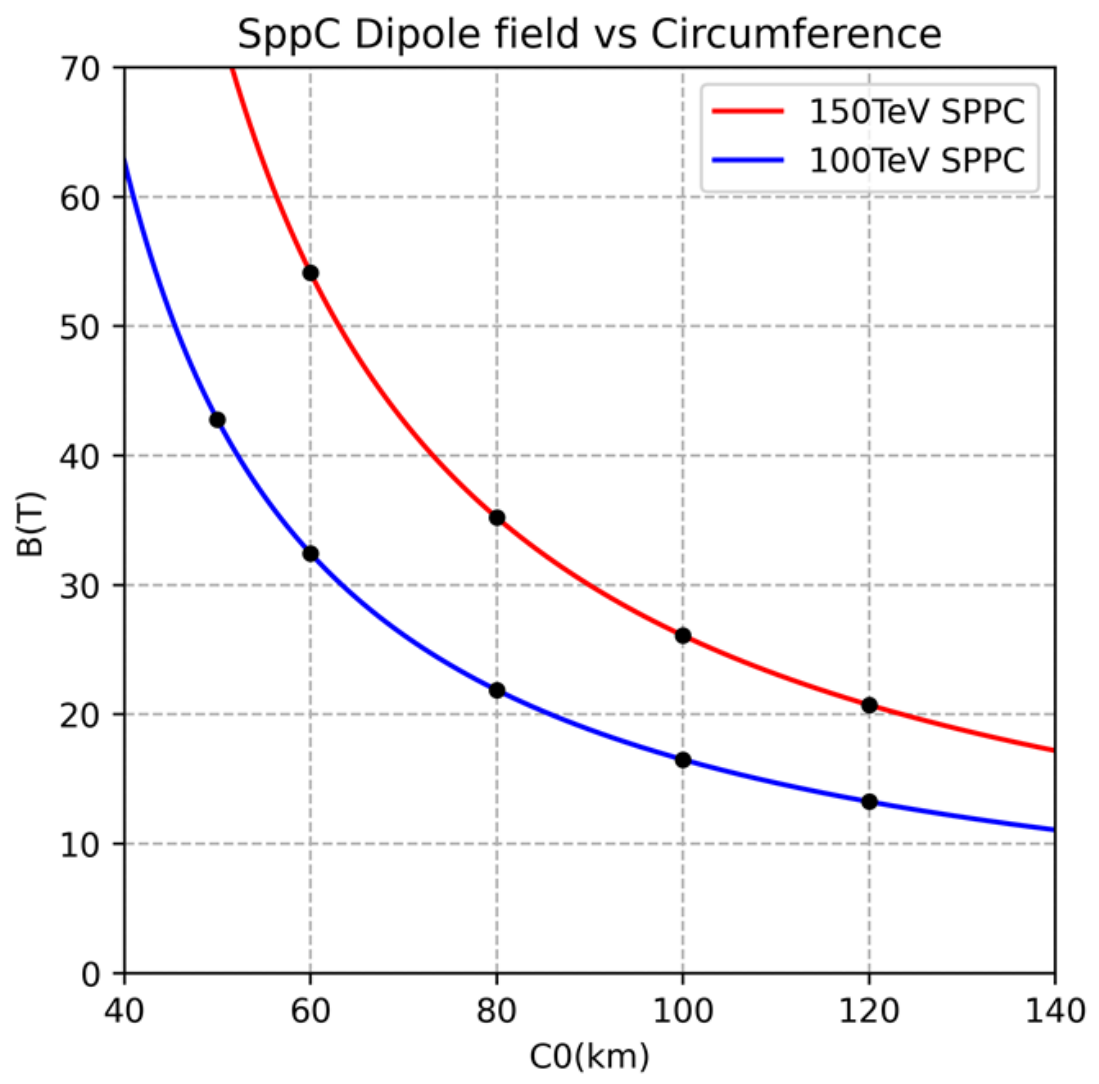

Fig. 17: The dipole strength of SPPC corresponding to the different C. M. energy vs machine circumference.

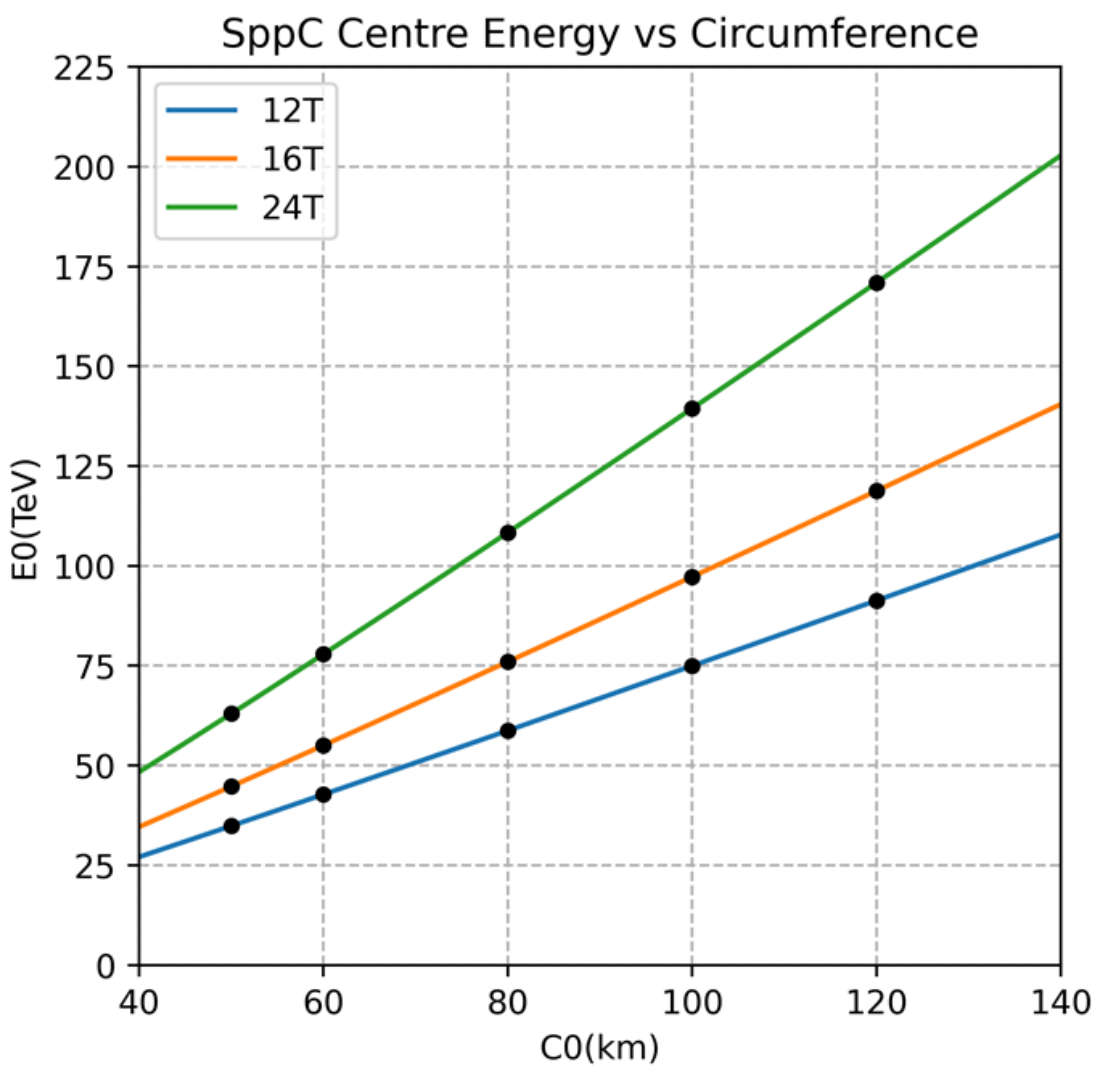

Fig. 18: The C. M. energy of SPPC corresponding to the different dipole strength vs machine circumference.

## 7. Summary

We have performed simplistic cost optimization studies for CEPC (Circular Electron Positron Collider) based on a rough cost model and the circumference choice of CEPC is reconsidered in a quantitative way. Higgs physics is the first goal of CEPC and hence the machine design is optimized for Higgs energy. If the total Higgs boson demand is about 1 million, a circumference of 80 km is a good choice, while 80 km and 100 km are almost the same if the total Higgs boson demand is more than 2 million. Combining the physics of Higgs and top quarks, 100 km is the best choice. Overall, 100 km circumference is a good choice for the CEPC, considering the compatibility of the machine and the future potential of ttbar, Z and SPPC.

## Reference:

[1] ATLAS Collaboration, Phys. Lett. B, 716: 1-29 (2012), arXiv:1207.7214[hep-ex].

[2] CMS Collaboration, Phys. Lett. B, 716: 30-61 (2012), arXiv:1207.7235[hep-ex].

[3]ATLAS and CMS Collaborations, Phys. Rev. Lett., 114: 19180 3 (2015), arXiv:1503.07589 [hep-ex]

[4] ATLAS Collaboration, Projections for Measurements of Higgs Boson Cross Section, Branching Ratios and Coupling Parameters with the ATLAS Detector at a HL-LHC, ATL-PHYS-PUB-2013-014(2013). https://cds.cern.ch/record/1611186 .

[5] ATLAS Collaboration, Search for the Standard Model Higgs and Z Boson Decays to J/ψ γ: HL-LHC Projections, ATL-PHYSPUB-2015-043(2015). http://cds.cern.ch/record/2054550.

[6] ATLAS Collaboration, Study of the Double Higgs Production Channel H(→ b)H(→ γγ) with the ATLAS Experiment at the HL-LHC, ATL-PHYS-PUB-2017-001, CERN, Geneva, 01020146 Chinese Physics C, 010201 Jan, 2017. https://cds.cern.ch/ record/2243387.

[7] The International Linear Collider Technical Design Report - Volume 3.I: Accelerator R&D in the Technical Design Phase, 2013.

[8] CLIC CDR, P. Lebrun et al, CERN 2012-005, arXiv :1209.2543.

[9] CEPC Accelerator Study Group, Snowmass 2021 White Paper AF3-CEPC, http://arxiv.org/abs/2203.09451.

[10] Future Circular Collider Study: The Lepton Collider (FCC-ee) Conceptual Design Report Volume 2, CERN-ACC-2018-0057, Eur. Phys. J. Spec. Top. 228 (2019) 261-623.

[11] Michael Benedikt, Alain Blondel, Patrick Janot, Michelangelo Mangano, Frank Zimmermann, Future Circular Colliders Succeeding the LHC, Nature Physics, 2020-04-06, Vol.16 (4), p.402-407

[12] Alain Blondel, Alex Chao, Weiren Chou, Jie Gao, Daniel Schulte, Kaoru Yokoya, Report of the ICFA Beam Dynamics Workshop: “Accelerators for a Higgs Factory: Linear vs. Circular” (HF2012), arXiv:1302.3318 [physics.acc-ph].

[13] CEPC-SPPC Preliminary Conceptual Design Report, March 2015, IHEP-CEPC-DR2015-01, http://cepc.ihep.ac.cn/preCDR/volume.html.

[14] CEPC-SPPC Progress Report (2015-2016), April 2017, IHEP-CEPC-DR-2017-01, http://cepc.ihep.ac.cn/Progress%20Report.pdf.

[15] CEPC Conceptual Design Report Volume I (Accelerator), August 2018, IHEP-CEPC-DR-2018-01, http://cepc.ihep.ac.cn/CEPC_CDR_Vol1_Accelerator.pdf.

[16] P. Lebrun and P. Garbincius, Assessing Risk in Costing High-energy Accelerators: from Existing Projects to the Future Linear Collider, Proc. IPAC2010 (2010) 3392.

[17] P. Lebrun, Costing High-energy Accelerator Systems, in proceedings of 4th RFTech Workshop, Annecy, France, 25–26 March 2013, http://lpsc.in2p3.fr/Indico/conferenceDisplay.py?confId=862.

[18] V. Shiltsev, A Phenomenological Cost Model for High Energy Particle Accelerators, JINST 9 T07002, 2014.

[19] V. Shiltsev and F. Zimmermann, Modern and Future Colliders, Reviews of Modern Physics, Vol. 93, 2021.

[20] J. Gao, Analytical Estimation of the Beam-beam Interaction Limited Dynamic Apertures and Lifetimes in e+e- Circular Colliders, Nucl. Instr. and methods A463（2001）p. 50-61.

[21] J. Gao, Emittance Growth and Beam Lifetime Limitations due to Beam-beam Effects in e+e- Storage Rings, Nucl. Instr. and methods A533（2004）p. 270-274.

[22] Pantaleo Raimondi, Dmitry Shatilov, Mikhail Zobov, Beam-beam Issues for Colliding Schemes with Large Piwinski Angle and Crabbed Waist, LNF-07/003 (IR), 2007.

[23] M. Zobov et al., Test of Crab-Waist Collisions at DAFNE Phi Factory, Phys. Rev. Lett. 104, 174801 (2010).

[24] Dou Wang, Jie Gao, et al., CEPC Partial Double Ring Scheme and Crab-waist Parameters, International Journal of Modern Physics A, Vol. 31, No. 33 (2016) 1644016.

[25] Alexander Wu Chao et al., Handbook of Accelerator Physics and Engineering (second edition), World Scientific, 2013, p. 312.

[26] Dou Wang, Jie Gao, et al., Optimization Parameter Design of a Circular e+e- Higgs Factory, Chinese Physics C, Vol. 37, No. 9 (2013) 097003.

[27] Dou Wang, Jie Gao, et al., 100 km CEPC Parameters and Lattice Design, International Journal of Modern Physics A, Vol. 32, No. 34 (2017)1746006.

[28] V. I. Telnov, Restriction on the Energy and Luminosity of e+e− Storage Rings due to Beamstrahlung, Phys. Rev. Letters 110,114801(2013).

[29] V. I. Telnov, Issues with Current Designs for e+e− and γγ Colliders, PoS (Photon 2013) 070.

[30] K. Oide, The Next Step for FCC-ee, FCC Week 2019, June 2019.

[31] K. Oide, Remaining Issues on the Design of Beam Optics for the FCC-ee Collider Rings, FCC Week 2020, Nov. 2020.

[32] K. Oide, FCC-ee Optics and Layout - Update and Plan, FCC Week 2021, June 2021.